\documentstyle[12pt,a4,epsfig]{article}
\begin{document}
\input psfig
\normalbaselineskip = 24 true pt
\normalbaselines
\addtolength{\topmargin}{-2cm}
\newcommand{\bm}{\bibitem}
\newcommand{\ud}{\bf}
\textheight=23cm
\renewcommand{\thefootnote}{\fnsymbol{footnote}}
\begin{center}
{\LARGE Nuclear breakup of $^8$B in a direct fragmentation model
\footnote{Work supported by GSI Darmstadt and BMFT. }}\\[1.0cm]
{\bf {R. Shyam$^{a}$\footnote{E-mail address: shyam@tnp.saha.ernet.in, associate
of Abdus Salam International Centre for Theoretical Physics, Trieste.}
and H. Lenske$^{b}$}}\\ 
$^{a}${\it Saha Institute of Nuclear Physics, 
Calcutta - 700 064, India}\\[0.2 cm]
$^{b}${\it Institut f\"ur Theoretische Physik, Universit\"at Giessen, 
D-35392 Giessen, Germany}\\ 
\end{center}
\renewcommand{\thefootnote}{\arabic{footnote}}
\begin{abstract}

We calculate the  cross sections of the elastic and inelastic breakup
modes for the inclusive breakup reaction $^{28}$Si($^8$B,$^7$Be)
at beam energies between 10 - 40 MeV/nucleon within a direct 
fragmentation model formulated in the framework of the post form
distorted-wave Born-approximation. In contrast to the case of the 
stable isotopes, the inelastic breakup mode is found to contribute
only up to 30$\%$ to the total breakup cross section, which is in
agreement with the recently measured experimental data. However,
the high energy tail of the energy spectra of $^7$Be fragment is
dominated by the inelastic breakup mode. The breakup amplitude is
found to be dominated by contributions from distances well beyond the 
nuclear surface. 
 
 
PACS NO. 25.60.Gc, 24.10.-i, 25.70.Mn  
\end{abstract}
\newpage
\section{Introduction}

Recently, there has been a lot of interest in the 
study of the proton drip-line nucleus $^8$B which is perhaps the
most likely candidate for having a proton halo structure ~\cite{Kita93,
Riis93, Riis94}, since its last proton has a binding energy of only 137
keV. Several measurements reported lately do seem to provide evidence in
favor of this possibility. For example, the electric quadrupole moment
of $^8$B is found to be twice as large as the value predicted by the
shell model, which can be explained with a single particle wave function
corresponding to a matter density of root mean square (rms) radius of
2.72 fm ~\cite{Mina92}. The observed narrow longitudinal momentum
($p_\parallel$) distribution of the $^7$Be fragment emitted in the
breakup reaction of 1.47 GeV/nucleon $^8$B on $^{12}C$ target has been 
interpreted in terms of a greatly extended proton
distribution in $^8$B ~\cite{Schw95}. The significantly enhanced
reaction cross sections of $^8$B measured at beam energies between
20 - 60 MeV/nucleon are shown to be consistent with the large matter radius
of $^8$B required to explain its quadrupole moment ~\cite{Warn95}.

Nevertheless, the existence of a proton halo in $^8$B is still an open
issue. Nakada and Otsuka ~\cite{Naka94} have shown that $(0 + 2)\hbar \omega$
shell model calculations can reproduce the measured large quadrupole 
moment of $^8$B without invoking a proton halo structure. The 
interaction cross sections measured by Tanihata et al. ~\cite{Tani88}
at 790 MeV/nucleon are consistent with a normal size of $^8$B.
The $p_\parallel$ distribution of the $^7$Be fragment emitted in the
breakup reaction of $^8$B at 41 MeV/nucleon has been found to be 
dependent on the target mass in Ref. ~\cite{Kell97} where it is  
argued that in contrast to the situation in the neutron halo
nuclei ~\cite{Kell95, Bane96}, the assumption of an unusually
extended spatial distribution is not necessary to explain the narrow
$p_\parallel$ distribution in case of $^8$B; the reaction mechanism
plays an important role here.

Breakup reactions in which the halo particle(s) is(are) removed from the 
projectile in the Coulomb and nuclear fields of the target nucleus, have 
played a significant role in probing the neutron halo structure in some 
light neutron drip-line nuclei (see e.g. ~\cite{Han95} for a recent review).
The enhanced total Coulomb breakup cross sections ~\cite{Blan91, Sack93},
narrow longitudinal momentum distributions of the heavy fragments ~\cite
{Kell95, Koba88, OrrN95}, and sharply forward peaked angular distributions of 
the valence neutron(s) ~\cite{Anne90, Anne94} are some of the pivotal
observations through which the neutron halo structure has been well 
manifested.
 
Apart from the $p_\parallel$ distributions of the $^7$Be fragment, some
data on the total cross section of the breakup reaction 
$^8$B + A $\rightarrow$ $^7$Be + X on low mass targets 
have also become available recently ~\cite{Nego96}. The theoretical studies 
reported so far have used either the Serber type ~\cite{Serb47} of models
~\cite{Nego96, Hans96} or the diffraction dissociation picture ~\cite{Henc96}
developed by Sitenko and co-workers ~\cite{Site90}. Both these approaches 
are essentially semi-classical in nature, hence a more microscopic
calculation within the quantum mechanical scattering theory is needed
to interpret the data properly. A proper understanding of the nuclear
breakup of $^8$B is also important in the context of the extraction of 
the astrophysical $S$-factor for the radiative capture reaction 
$^7$Be(p,$\gamma$)$^8$B from the Coulomb dissociation of $^8$B
~\cite{Baur97,Kiku97}.

In this paper, we present calculations of the cross sections for 
the breakup reaction $^8$B $+$ $^{28}$Si $\rightarrow$ $^7$Be $+$ X
within a direct fragmentation model (DFM), which is formulated in
the framework of the post form distorted wave Born approximation
(PFDWBA) ~\cite{Shya92, Shya85, Baur84}. As the target nucleus involved
in this reaction is very light, we shall consider only the nuclear
breakup process. However, the Coulomb breakup can also be calculated in
this theory on the same footing (see e.g. ~\cite{Shya92}). In the next
section we present the details of our formalism. In section 3, our
results are presented and discussed. The conclusions of our work are
described in section 4.

\section{Formalism}

The nuclear breakup cross section consists of two components: the
elastic breakup (ELB) (also known as "diffraction dissociation") where
X corresponds to the target nucleus A in its ground state and proton, and
inelastic breakup (INELB) (also known as "diffraction stripping") where
X can be any other channel of the A $+$ p system.
The triple differential cross section for the elastic
breakup reaction $a + A \rightarrow c + x + A$ (e.g for our case, 
$a$ = $^8$B, $c$ = $^7$Be, $x$ = p), is defined as 
\begin{eqnarray}
\frac{d^3\sigma^{ELB}(a,c)}{d\Omega_c dE_c d\Omega_x} & = & 2\pi
            \frac{\mu_a \mu_c \mu_x} {(2\pi \hbar)^6} \frac{q_c q_x}{q_a}
            \sum_{\ell m_{\ell}} \mid \beta_{\ell m_{\ell}}^{ELB} \mid^2,
\end{eqnarray}
where the transition amplitude $\beta_{\ell m_{\ell}}$ is given by
\begin{eqnarray}
\sqrt{(2\ell+1)}\beta_{\ell m_\ell}^{ELB} & = & \int dr_{cx} dR_i
                              \chi^{(-)*}({\bf q}_c, {\bf R}_f)
                              \chi^{(-)*}({\bf q}_x, {\bf R}_x) \nonumber \\ 
                     &   & \times   V_{cx}({\bf r}_{cx}) u_{\ell}(r_{cx})
                               y_{\ell m_\ell}(\hat r_{cx})
                               \chi^{(+)}({\bf q}_a,{\bf R}_i).
\end{eqnarray}
In Eqs. (1) and (2), $\ell$ is the orbital angular momentum for relative
$c$ + $x$ system and $y_{\ell m_\ell}$ are the spherical harmonics.
$V_{cx}$ represents the interaction between constituents $c$ and $x$  
while $u_\ell$ is the wave function for their relative motion
in the projectile ground state. 
${\bf q}_a$ ($\mu_a$), ${\bf q}_c$ ($\mu_c$) and ${\bf q}_x$ ($\mu_x$) 
are the momenta (reduced masses) of the particles $a$, $c$ and $x$
respectively. $\chi$'s denote the scattering wave functions which are
generated by the appropriate optical potentials in respective channels.
The system of coordinates used are the same as that given in Ref.
~\cite{Shya85}.

The transition amplitude is a six dimensional integral. By making 
a zero range approximation (ZRA) this integral is reduced to three
dimensions ~\cite{Satc83}, although its calculation is still a major
problem as it involves a product of three scattering waves which converge
very slowly. In the ZRA the details of the internal structure of the
projectile appear in the amplitude only through an overall
normalization constant and the values of $\ell$ other than zero are
necessarily excluded. Because of the relative p-state between
$^7$Be and the proton in the ground state of $^8$B the ZRA, 
therefore, is not suitable for this case.  However,
we introduce a constant range approximation (CRA) which reduces the
integral in Eq. (2) to three dimensions and at the same time allows
the non-zero values of $\ell$ to enter in the calculations. We 
assume that the breakup reaction is strongly peripheral and that only
those configurations where (1) the proton is in the collinear position
between the target nucleus and $^7$Be and (2) the relative separation
between proton and $^7$Be is constant (say $d$ $fm$), contribute to the
transition amplitude. We can then write 
\begin{eqnarray}
V({\bf r}_{cx})u_\ell(r_{cx})y_{\ell m_\ell}(\hat {r}_{cx}) & = & D_a
                      \delta({\bf r}_{cx} - d \frac{{\bf R}_i}{R_i}) 
                      y_{\ell m_\ell}(\hat {R}_i),
\end{eqnarray}
where $d$ is taken to be the distance between origin and the maximum in
the $^8$B ground state wave function. Approximations similar to the CRA
have been used earlier by P\"uhlhoffer et al. ~\cite{Puhl70} and Kubo and
Hirata ~\cite{Kubo72} to study the $\alpha$ transfer reactions induced by 
$^6$Li and $^7$Li projectiles within the DWBA. These authors have found a
reasonable agreement between the calculated and measured angular distributions
particularly for momentum matched transitions. Furthermore, as long
as the reaction is not sensitive to the smaller distances, the results
obtained with Eq. (3) are in agreement with those of the full finite range
calculations. 

Substituting Eq. (3) in Eq. (2), making partial wave expansion for 
$\chi^{(-)*}({\bf q}_x, {\bf R}_x)$
and using Eq. (4.13) of ~\cite {Brin93}, we get
\begin{eqnarray}
\beta_{\ell m_\ell}^{ELB} & = & \sqrt{4\pi} D_a \sum_{\ell_x m_x}
                        \sum_{L M_L} \imath^{-\ell_x} (-)^{m_x}
        \sqrt{\frac{2\ell_x+1}{2L+1}} y_{\ell_x m_x}(\hat {q}_x) \nonumber \\
                    &   & \times <\ell_x -m_x \ell m_\ell \mid L M_L>
                        <\ell_x 0 \ell 0 \mid L 0> 
                        \tilde {\beta}_{\ell m_\ell}^{ELB},
\end{eqnarray}
where the reduced amplitude $\tilde {\beta}_{\ell m_\ell}^{ELB}$ is given by
\begin{eqnarray}
\tilde {\beta}_{\ell m_\ell}^{ELB} & = & \int dR_i
                        \chi^{(-)*}({\bf q}_c, \alpha_1 {\bf R}_i)
                     \frac{\chi_{\ell_x}(q_x, \alpha_2 R_i)}{\alpha_2 q_x R_i}
                     y_{L M_L}(\hat {R}_i) \chi^{(+)}({\bf q}_a,{\bf R}_i).
\end{eqnarray}
In Eq. (5) we have  
\begin{eqnarray} 
\alpha_1 & = &\frac{m_A}{m_A+m_x} + (1 - \frac{m_A}{m_A+m_x}\frac{m_c}{m_a})
                            \frac{d}{R_i},\\
\alpha_2 & = & 1 - \frac{m_c}{m_a}\frac{d}{R_i},
\end{eqnarray}
and $\chi_{\ell_x}$ is the radial part of the wave function
$\chi^{(-)*}({\bf q}_x, {\bf R}_x)$. Eq. (5) is similar to the
amplitude obtained with the ZRA and
can be evaluated by using the method described in e.g.~\cite{Baur84}.
The expressions for the zero range amplitudes are retrieved from 
Eq. (4) - Eq. (6) by assuming $\ell$ and $d$ equal to zero.

The transition amplitude for the inelastic breakup reaction 
$a + A \rightarrow c + C$, where $C$ is some final state of the
system $A + x$, is given by ~\cite{Shya85}
\begin{eqnarray}
\sqrt{(2\ell+1)}\beta_{\ell m_\ell}^{INELB} & = & \int dr_{cx} dR_i
                              \chi^{(-)*}({\bf q}_c, {\bf R}_f)
                              \chi^{C}({\bf q}_x, {\bf R}_x) \nonumber \\
                       &   & \times V_{cx}({\bf r}_{cx}) u_{\ell}(r_{cx})
                               y_{\ell m_\ell}(\hat r_{cx})
                               \chi^{(+)}({\bf q}_a,{\bf R}_i),
\end{eqnarray}
where $\chi^{C}({\bf q}_x, {\bf R}_x)$ is the form factor,
which is obtained by taking the overlap of the wave function for the channel
$C$ (which incorporates all the possible reactions initiated by the 
interaction between nuclei $A$ and $x$), with the wave function describing the 
internal states of the target and projectile nuclei. Using Eq. (3) 
and other steps as described above, an expression similar to Eq. (4)
can be derived for $\beta_{\ell m_\ell}^{INELB}$ with the reduced
amplitude given by 
\begin{eqnarray}
\tilde {\beta}_{\ell m_\ell}^{INELB} & = & \int dR_i
                        \chi^{(-)*}({\bf q}_c, \alpha_1 {\bf R}_i)
                     \frac{\chi_{\ell_x}^C(q_x, \alpha_2 R_i)}{\alpha_2 q_x R_i}
                     y_{L M_L}(\hat {R}_i) \chi^{(+)}({\bf q}_a,{\bf R}_i),
\end{eqnarray}
where $\chi_{\ell_x}^{C}$ is the radial part of the form factor. Its
calculation simplifies greatly if we introduce the so called "surface
approximation" and write $\chi_{\ell_x}^{C}$ in terms of its asymptotic form
\begin{eqnarray} 
\chi_{\ell_x}^{C}(q_x, R_i) & = & \frac{1}{2} S_{\ell_x,C}
                                  H_{\ell_x}^{(+)}(q_x,R_i),
\end{eqnarray}  
where $H_{\ell_x}^{(+)} = G_{\ell_x} + \imath F_{\ell_x}$, with $F_{\ell_x}$
and $G_{\ell_x}$ being the regular and irregular Coulomb wave functions. 
This equation can be rewritten in terms of the elastic scattering 
wave function $\chi_{\ell_x}$ ( see Eq. (5)) as
\begin{eqnarray} 
\chi_{\ell_x}^{C}(q_x, R_i) & = & \frac{S_{\ell_{x},C}}{S_{\ell_{x},\ell_{x}}-1}
                        (\chi_{\ell_x}(q_x,R_i) - F_{\ell_x}(q_xR_i)),
\end{eqnarray}
where $S_{\ell_x,\ell_x}$ are the $S$ matrix elements for
the elastic channel corresponding to the angular momentum $\ell_x$. The
validity of the surface approximation has been tested by Kasano and Ichimura
~\cite{Kasa82} who found it to be well fulfilled even for the deuteron. 
We use Eq. (11) for the form factor $\chi_{\ell_x}^C$ also
in the interior region in Eq. (9), which is not expected to be a serious
approximation as this region contributes very little to the whole DWBA
integral. In order to calculate the inelastic breakup cross section
one has to sum over all the channels $C \neq \ell_x$, which can be easily
done by using the unitarity of the $S$ matrix as all the
dependence on channel $C$ in the transition amplitude rests in the 
$S$ matrix $S_{\ell_x,C}$
\begin{eqnarray}
\sum_{C \neq \ell_x} \mid S_{\ell_x,C} \mid^2 & = & 1
                                   - \mid S_{\ell_x,\ell_x}\mid^2 
\end{eqnarray}
Thus the inelastic breakup cross section can be written as 
\begin{eqnarray}
\frac{d^3\sigma^{INELB}(a,c)}{d\Omega_c dE_c d\Omega_x} & = & 2\pi
            \frac{\mu_a \mu_c \mu_x} {(2\pi \hbar)^6} \frac{q_c q_x}{q_a}
   \sum_{\ell m_{\ell}}(\sigma_{\ell_x}^{REAC}/\sigma_{\ell_x}^{ELAS})
   \mid {\beta}_{\ell m_{\ell}}^{ELB} 
   - {\beta}_{\ell m_{\ell}}^{0} \mid^2,
\end{eqnarray} 
where ${\beta}_{\ell m_{\ell}}^{0}$ is the same as the amplitude defined in
Eq. (4) with the wave function $\chi_{\ell_x}$ replaced by the regular 
Coulomb function. The partial reaction and elastic cross sections
$\sigma_{\ell_x}^{REAC}$ and $\sigma_{\ell_x}^{ELAS}$ are related to
the scattering matrix elements $S_{\ell_x,\ell_x}$ by their usual definitions.
It may be noted that the quantities required to calculate the inelastic
breakup are the same as those already calculated in the case of elastic
breakup.

In order to study the impact parameter dependence of the breakup
cross section, we define the "probability of breakup"
$T_{\ell_a}^{b-up(a,c)}$ as
\begin{eqnarray}
\sigma_{total}^{b-up(a,c)} & = & \int dE_c d\Omega_c d\Omega_x
                            \frac{d^3\sigma(a,c)}{dE_c d\Omega_c d\Omega_x}\\
                        & = & (\pi/q_a^2)\sum_{\ell_a}
                              (2\ell_a + 1)T_{\ell_a}^{b-up(a,c)}
\end{eqnarray}
In Eq. (14), $d^3\sigma(a,c)/dE_c d\Omega_c d\Omega_x$ is the sum of 
the elastic and inelastic breakup cross sections given by Eqs. (1) and (13). 

\section{Results and discussion}

The optical potentials in the entrance and outgoing channels and the
constants $D_a$ and $d$ are required as input in our numerical calculations.
Although some elastic scattering data for the $^8$B, $^7$Be + $^{12}$C
systems at the beam energy of 40 MeV/nucleon are available ~\cite{Peci95},
the usual optical model fits to them is unfortunately not 
reported. Unless otherwise stated, our calculations have been performed 
with the following set of optical potentials, $V_R$ = 123.0 MeV, $r_V$ = 
0.75 fm, $a_V$ = 0.80 fm, $W$ = 65.0 MeV, $r_W$ = 0.78 fm, $a_W$ = 0.80 fm,
with real and imaginary volume Woods-Saxon terms. This potential, which  
is similar to that used in the recent analysis of $^9$Li and $^{11}$Li
elastic scattering ~\cite{Zaha96}, has been used for both $^8$B and $^7$Be. 
The $(A^{1/3} + a^{1/3})$ convention was followed to get the radii from the
radius parameters. The global Becchetti-Greenlees parameterization
~\cite{Becc69} was used for the proton-target potential.

The constants $D_a$ (see e.g.~\cite {Kubo72}) and $d$ have been determined with
a $^8$B ground state wave function obtained by assuming it to be a pure
$0p_{3/2}$ proton single particle state, with separation
energy 0.137 MeV, calculated in a central Woods-Saxon potential of
geometry, $r_0$ = 1.54 fm, and $a_0$ = 0.52 fm ~\cite{Kim87}. This gives
$D_a$ = -39.0 MeV fm$^{3/2}$ and $d$ = 1.8 fm, which has been used
in all the calculations described in this paper.
The radius and diffuseness parameters used by Barker ~\cite{Bark80}
($r_0$ = 1.25 fm, and $a_0$ = 0.65 fm), and Esbensen and Bertsch
($r_0$ = 1.25 fm, and $a_0$ = 0.52 fm) ~\cite{Esbe96} lead to $D_a$ = -40.4
and -41.7 MeV fm$^{3/2}$ and $d$ = 1.8 and 1.7 fm respectively.
On the other hand, with more elaborate RPA models for the $^7$Be + p
overlap wave function ~\cite{Schw95}, the values of $D_a$ and $d$ are found
to be -38.0 MeV fm$^{3/2}$ and 1.8 fm respectively. Hence, the   
constants $D_a$ and $d$ do not show any marked dependence on the 
nuclear structure model of $^8$B. For the sake of comparison
with other approaches the simplified potential model as used by us
seems to be adequate at this stage of the present theory.     

Fig. 1 shows the calculated elastic (dotted line), inelastic (dashed line)
and total (solid line) breakup
cross sections for the $^8$B $+$ $^{28}$Si $\rightarrow$ $^7$Be $+$ X
reaction as a function of beam energy together with the data taken from
~\cite{Nego96}. We see that the measured total breakup cross sections
are well reproduced by our calculations although the contributions of the 
elastic and inelastic breakup modes are slightly over- and under-predicted
respectively. The breakup cross sections decrease with beam energy up to
20 MeV/A and after that they are almost constant (although the inelastic
breakup mode still shows a tendency of decreasing somewhat). 
The nuclear breakup cross sections of $^8$B as reported in ~\cite{Henc96,Nego96}
show a similar type of energy dependence in this beam energy regime although
their increase below 20 MeV/nucleon is less pronounced than that seen
in Fig. 1. Clearly, more measurements are needed to clarify this point.
 
A striking feature of the results shown in Fig. 1 is that the 
contribution of the inelastic breakup mode to the total breakup cross
section is limited only to about 30$\%$, which is in agreement with the 
experimental data ~\cite{Nego96}. This is in marked contrast
to the situation in stable nuclei where this mode of breakup makes
up about 75 - 80 $\%$ of the total $(a,c)$ breakup cross section 
(see e.g. ~\cite{Baur84}). To understand this difference, we show in Fig. 2
the breakup probability ($T_\ell$) (defined by Eq. (14)) for the 
reaction $^{28}$Si($^8$B, $^7$Be) as a function of the impact parameter
$b$ ($b = (\ell_d + 1/2)/q_a$). It can be seen that $T_\ell$ 
peaks at about 8 fm, which is in remarkable agreement with that
obtained in Ref. ~\cite{Nego96} from the semi-classical arguments. This
is quite large in comparison to the sum of the radii ($R_s$) of $^8$B and
$^{28}$Si ($\simeq$ 6 fm). Furthermore, most of the contribution
to $T_\ell$ comes from the impact parameters beyond 8 fm,
while those from distances $<$ $R_s$ are strongly suppressed. This
clearly shows that the breakup of $^8$B takes place far away from the
nuclear surface which reduces the probability of the inelastic breakup
process; large impact parameters favor the elastic breakup mode.
In contrast, for stable isotopes, the breakup probability peaks around
$R_s$ (where the inelastic breakup mode is maximum) and the drop from
the peak value for $b > R_s$ is much faster than that seen in
Fig. 2 ~\cite{Baur84, Baur80}. This, explains to some extent the
difference in the nature of the inelastic breakup cross section of
$^8$B as compared to that of the stable isotopes. The
fact that the breakup of $^8$B is dominated by contributions coming
from a large range of impact parameters $>R_s$, is in agreement with
the observation made earlier in the case of neutron halo nuclei
$^{11}$Li and $^{11}$Be ~\cite{Shya92}. This could provide an indirect
evidence for a larger spatial extension of the proton in the ground
state of $^8$B. It should, however, be stressed that a $0p_{3/2}$
configuration for the p - $^7$Be relative motion already leads to a 
$^8$B matter density with a larger rms radius ~\cite{Naka94,Brow96}.
Nevertheless, in the present calculations the $^8$B structure input 
largely affects only the absolute magnitudes of the cross sections  
(through the constant $D_a$); the peak position in Fig. 2 is mostly
decided by the reaction dynamics. It is possible, in principle, to include
other components (eg. $p_{1/2}$, $f_{7/2}$) in the $^8$B wave function
within this formalism. However, as the $0p_{3/2}$ component carries by far
the largest spectroscopic weight ($>$ 90 $\%$), this is unlikely to alter
our results much. 

As a side remark, we point out that the relative contributions of the
elastic and inelastic breakup modes are independent of the uncertainties in 
the values of $D_a$ and $d$ as the same constants enter in all the cross
sections. 
 
In Fig. 3, we show the contributions of elastic and inelastic breakup
modes to the energy distribution of the $^7$Be fragment at the beam 
energy of 30 MeV/nucleon. One notes that while 
the elastic breakup mode dominates in the peak region, its contribution 
is very weak towards the high energy end (where the proton energy
$E_p \rightarrow 0$); total cross section is made up mostly of
the inelastic breakup mode in this region. The threshold behavior of
the breakup cross section can be easily understood from that of
the phase-shift $\delta_p$ of the scattering of the proton from the target.
It can be shown that ~\cite{Baur80} in the limit $E_p \rightarrow 0$, the
elastic and inelastic breakup cross sections are proportional
to $q_p^{2\ell_p+1}$ and $R^{2\ell_p+1}$ (where $R$ is the
radius of the nuclear potential) respectively.
Therefore, in this limit the elastic breakup cross section
tends to zero even for the $s-$ wave A + p interaction while the inelastic 
one to a finite value, which explains the observation made in Fig. 3.
It would be very interesting therefore, to perform measurements for
the energy spectra of the fragment $^7$Be to confirm the 
$E_p \rightarrow 0$ behavior. It may help in fixing the absolute
magnitude of the inelastic breakup cross section in the breakup experiments. 

\section{Conclusions}

To conclude, we have presented for the first time a fully quantum mechanical
calculation of the elastic and inelastic modes of the nuclear breakup
of $^8$B on the Si target. We employed a direct fragmentation model
which is formulated within the framework of the post form distorted wave
Born-approximation. This is a definite improvement over the semi-classical
models of the breakup reactions used so far for this purpose. 
We obtain a good overall description of the experimental data measured
recently. The inelastic breakup mode is found to contribute only up to 30$\%$
to the total breakup cross section for the $^{28}$Si($^8$B,$^7$Be) reaction,
which is in contrast to the breakup of the stable isotopes. Most of the
contributions to the breakup cross section come from the distances far
beyond the nuclear surface which favors the elastic breakup mode. 

The energy spectra of the $^7$Be fragment is dominated by the
inelastic breakup mode towards the high energy end where the proton
energy goes to zero. This observation which is beyond the scope of the 
semi-classical models, is a natural outcome of the PFDWBA theory of the
breakup reactions and it should be verified experimentally.

We must stress that the lack of the precise knowledge about the parameters
of the optical potential for the $^8$B,$^7$Be $+$ $^{28}$Si system is a
potential source of uncertainty in our calculations.
Therefore, the analysis of the existing data (taken at around 40 MeV/nucleon)
for this system ~\cite{Peci95} in terms of the conventional optical model
would be extremely useful. Moreover, similar studies at other beam energies
are also clearly needed. The present calculations are not very sensitive to the
nuclear structure models of $^8$B. The direct fragmentation
model should be improved further so that more detail of the projectile wave
function can enter in the calculations explicitly.  
Once such an extended reaction description is available, the
use of more elaborate nuclear structure models of the projectile
will be meaningful. It may then become possible to use the breakup data to 
distinguish the wave functions of $^8$B obtained from a static potential model  
from those calculated from dynamical approaches ~\cite{Schw95} where
the deformation of this nucleus is taken into account.

Useful discussions with Hans Geissel is gratefully acknowledged. Thanks are
also due to Ulrich Mosel for his very kind hospitality to one of us (RS) in
the University of Giessen.

\newpage
\begin{center}{\bf Figure Captions} \end{center}
\begin{itemize}
\item[Fig. 1.] Total cross section for the breakup reaction 
of $^8$B $+$ $^{28}$Si $\rightarrow$ $^7$Be $+$ X as a function of the
beam energy. The contributions of the elastic and inelastic 
breakup modes are shown by dotted and dashed lines respectively while 
their sum is depicted by the solid line. The experimental data for the 
total (solid circles), elastic (solid squares) and inelastic (open circles)
breakup cross sections are taken from ~\cite{Nego96}. 

\item[Fig. 2.] Breakup probability ($T_l$) (as defined by Eq. (14)) for the 
reaction studied in Fig. 1 at the beam energy of 40 MeV/nucleon as a
function of the impact parameter.

\item[Fig. 3.] Energy distribution of the $^7$Be fragment emitted in the
breakup of $^8$B on $^{28}$Si target at the beam energy of 30 MeV/nucleon. The
solid, dashed and dotted lines have the same meaning as in Fig. 1.
\end{itemize}
     
\begin{figure}
\begin{center}
\mbox{\epsfig{file=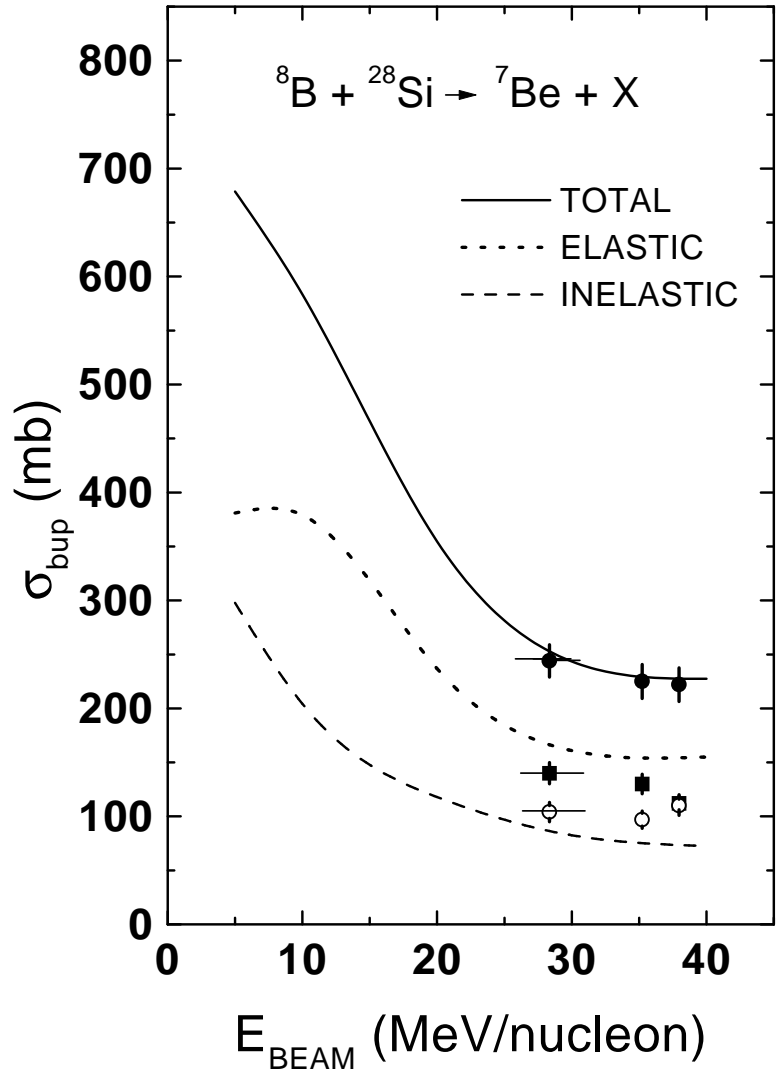,height=17.0cm}}
\end{center}
\caption{
Total cross section for the breakup reaction 
of $^8$B $+$ $^{28}$Si $\rightarrow$ $^7$Be $+$ X as a function of the
beam energy. The contributions of the elastic and inelastic 
breakup modes are shown by dotted and dashed lines respectively while 
their sum is depicted by the solid line. The experimental data for the 
total (solid circles), elastic (solid squares) and inelastic (open circles)
breakup cross sections are taken from ~\cite{Nego96}.} 
\label{fig:figa}
\end{figure}

\begin{figure}
\begin{center}
\mbox{\epsfig{file=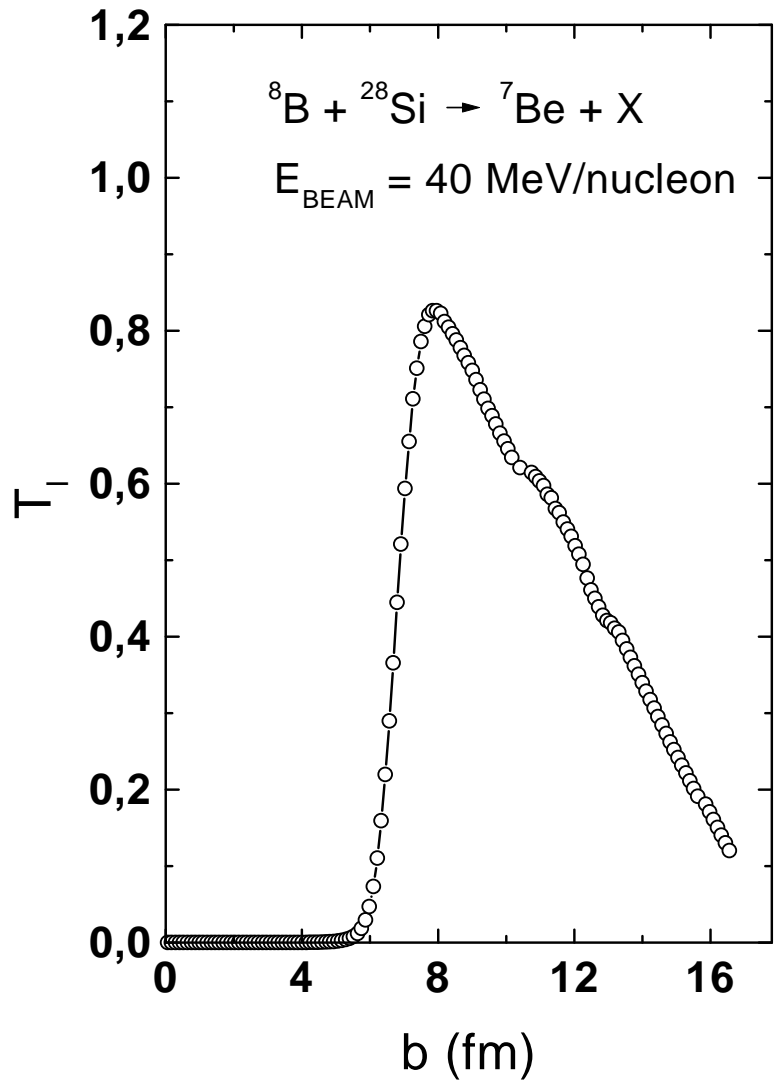,height=17.0cm}}
\end{center}
\caption{
Breakup probability ($T_l$) (as defined by Eq. (14)) for the 
reaction studied in Fig. 1 at the beam energy of 40 MeV/nucleon as a
function of the impact parameter.}     
\label{fig:figb}
\end{figure}

\begin{figure}
\begin{center}
\mbox{\epsfig{file=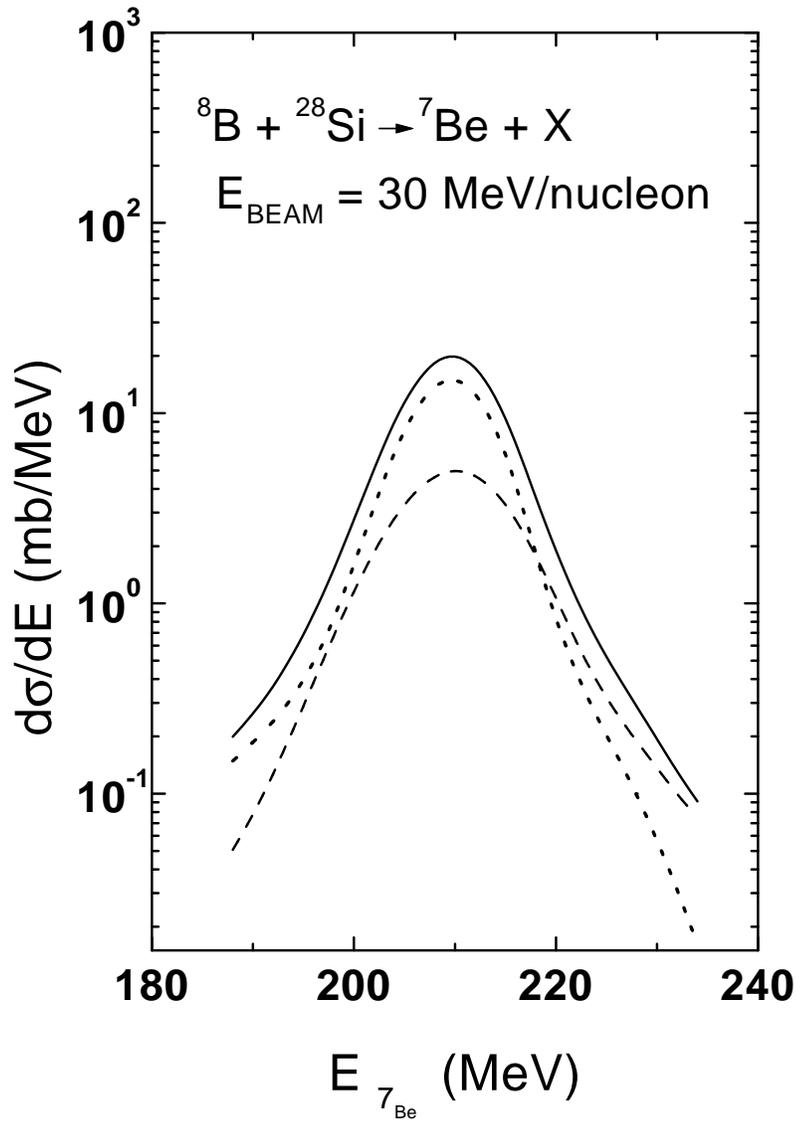,height=17.0cm}}
\end{center}
\caption{
Energy distribution of the $^7$Be fragment emitted in the
breakup of $^8$B on $^{28}$Si target at the beam energy of 30 MeV/nucleon. The
solid, dashed and dotted lines have the same meaning as in Fig. 1.}
\label{fig:figc}
\end{figure}
\end{document}